\documentclass[a4paper,aps,floatfix,twocolumn]{revtex4}

\usepackage[latin1]{inputenc}
\usepackage[OT1]{fontenc}

\usepackage{amsmath}
\usepackage{amssymb,amsfonts}
\usepackage{array}
\usepackage{dcolumn}

\usepackage{calc}

\usepackage{graphicx}
\usepackage{color}  

\begin{document}
\newlength{\colwid}
\setlength{\colwid}{\columnwidth}
\newcommand{\ord}[1]{\ensuremath{\mathcal{O}\bigl(#1\bigr)}}
\newcommand{\vect}{\ensuremath{\mathbf}}
\newcommand{\E}{\vect{E}}
\newcommand{\eps}{\epsilon}
\newcommand{\R}{\vect{r}}
\renewcommand{\div}{\mathrm{div}\,}

\title{Monte Carlo algorithms for charged lattice gases}
\author{L. Levrel and A. C. Maggs}
\affiliation{Laboratoire de Physico-Chimie Théorique, UMR CNRS-ÉSPCI
  7083, 10 rue Vauquelin, F-75231 Paris Cedex 05, France}
\date{\today}
\begin{abstract}
  We consider Monte Carlo algorithms for the simulation of charged
  lattice gases with purely local dynamics. We study the mobility of
  particles as a function of temperature and show that the poor
  mobility of particles at low temperatures is due to ``trails'' or
  ``strings'' left behind after particle motion. We introduce modified
  updates which substantially improve the efficiency of the algorithm
  in this regime.
\end{abstract}
\maketitle

\section{Introduction}

The properties of many condensed matter systems can not be understood
without considering the Coulombic interaction.  DNA, proteins,
polyelectrolytes, colloids and even water are all structured by
electrostatics, which must be reproduced faithfully in any numerical
study. Unfortunately, the simulation of the electrostatic interactions
is difficult; due to the slow decay of the potential in $1/r$ one can
not truncate the interaction \cite{frenkel} as is often done with
other molecular interactions. Most working codes now use a variant of
the Ewald sum to account for the interaction between periodic images
of the basic simulation cell. As a consequence the time need to
evaluate the electrostatic interaction can dominate in the simulation
of charged systems.

In a Monte Carlo simulation with the Ewald method, the motion of a
single charge requires summing its interactions with the $N-1$ other
charges and their periodic images, resulting in a $\ord{N^2}$
computational cost per sweep. This becomes impractical when $N$ is
large. In simulations with explicit modeling of all charges $N>10^4$
is commonly required. In molecular dynamics the situation is better:
when all particles are moved simultaneously, better CPU time scalings
are possible ranging from $\ord{N}$ for multigrid algorithms
\cite{multigrid} to $\ord{N^{3/2}}$ for an optimized Ewald summation
\cite{ewald32}. However these molecular dynamics codes are complex to
implement.

The unfavorable complexity of conventional Monte Carlo methods
originates in the use of the electrostatic potential $\Phi$, which is
the solution to Poisson's equation
\begin{equation*}
  \nabla^2 \Phi = -\rho/\eps.
\end{equation*}
This equation has a unique solution for given charge distribution and
boundary conditions. When a charge is moved, the new solution for $\Phi$
is computed and the interaction energy $q_i\Phi(\R_i)$ of the moved
charge with \emph{all} other charges $q_i$ in the system changes. The
electrostatic interaction implemented in this way is
\emph{instantaneous}. Note however \cite{PRL88}, the thermodynamical
study of charged systems does not require instantaneous Coulombic
interactions: The free energy is also correctly sampled when only
Gauss's law
\begin{equation*}
  \div\E=\rho/\eps
\end{equation*}
is imposed on the electric field.  The fact that solutions to Gauss's
law are \emph{not} unique results in an extra flexibility which allows
one to implement a purely local Monte Carlo scheme for the simulation
of systems with electrostatic interactions. The computation effort is
reduced to $\ord{N}$ per Monte Carlo sweep. The disadvantage of the
algorithm is that it requires a grid to discretize the electrostatic
degrees of freedom, however this is also true of multigrid and Fourier
methods used for molecular dynamics.

The final efficiency of the Monte Carlo algorithm depends on the
number of sweeps required to sample independent configurations
which in turn is a function of the particle mobility resulting from
the Monte Carlo dynamics. Highly mobile charges enable one to generate
independent configurations rapidly; if charges were to become
``trapped'' or ``localized'' due to their interaction with the field
it could prevent the generation of uncorrelated samples. Monitoring
the acceptance rate of particle updates may only give partial
information in that on the total efficiency of an algorithm. For
instance trapped particles could move locally (resulting in a good
acceptance rate) without being able to explore all of space.

In this article we perform a detailed study of the charge mobility
$\mu$ in local Monte Carlo algorithms in order to compare efficiencies
of various implementations. Firstly we develop a technique to measure
$\mu$ by relating the mobility to the dynamics of the average electric
field, $\bar{\E}$.  The mobility will be studied as a function of
temperature. With our previous implementation, $\mu$ drops
dramatically at low temperatures, becoming unmeasurable for parameters
which are needed to study typical materials: For instance monovalent
ions in water at room temperature where $\eps=78$, $T=300~\text{K}$,
$a=1~\text{Å}$ with $a$ the mesh size. The drop in efficiency
originates in the constrained dynamics of the electric field, leading
to the generation of ``trails'' or ``strings'' which trap particles at
low temperature and suppress their mobility.

We will explore ways of reducing this trapping. The update law
introduced previously \cite{PRL88} for particle motion is not the only
way one can move a charge. Even if each charge update must be
accompanied by \emph{some} field update, the latter is only loosely
constrained.
Duncan, Sedgewick, and Coalson recently used this fact to introduce
\cite{coalson} a better  particle-plaquette update.
We will present several field updating schemes
leading to less trapping. These schemes are very flexible in
that they have a freely adjustable ``spreading'' parameter $w$, upon
which their effects and their computational complexity depend. Schemes
which use a larger spreading parameter are more time-consuming but
lead to much larger efficiencies at physically interesting
temperatures.

We have already shown \cite{JCP120b} that off-lattice implementations
of the algorithm (using continuous interpolation of charges with
splines) do not suffer the mobility drop that we discuss in this
paper. Rather, our present work is motivated by the existence of a
wide spectrum of interesting and useful lattice models. For example,
it is known \cite{panaPRL83,panaJCP112,panaJCP118} that finely
discretized lattice fluids exhibit the same critical behavior as
continuum fluids. Another example is the bond fluctuation model for
polymers \cite{kremer,binder} which is rather easily generalized
to study charged polymers, or polyelectrolytes. All these models already
use ``spread'' or extended particles where the hard cores of the particles
span several lattice sites in order to reduce lattice artefacts to an
acceptable level.

We will begin (Sec.~\ref{sec:method}) with a description of the theoretical
basis and implementation of local Monte Carlo algorithms with
electrostatics. In Sec.~\ref{sec:mobility} we show how to measure
the mobility of charges and apply the method to the simplest
algorithm. We interpret the behavior of the acceptance rate and
mobility as a function of temperature (Sec.~\ref{sec:interpret}),
introducing the concept of field trails or strings. We show how to
increase particle mobility in Secs.~\ref{sec:spr},~\ref{sec:flum}.
Finally we will present the CPU time for representative simulations
and give the reader an estimate of optimal parameters.

\newcommand{\Elk}{E_{AB}}
\newcommand{\Etr}{\vect{E}_\perp}
\newcommand{\Q}{\vect{Q}}
\newcommand{\J}{\vect{J}}
\renewcommand{\j}{\vect{j}}
\renewcommand{\k}{\vect{k}}
\newcommand{\V}{\mathcal{V}}
\newcommand{\grad}{\mathrm{grad}\,}
\newcommand{\cross}{\ensuremath{\negmedspace\times\negmedspace}}
\newcommand{\rot}{\mathrm{curl}\,}
\def\D{\mathrm{d}}
\newcommand{\DR}{\ensuremath{\:\D³r}}
\newcommand{\DRi}{\ensuremath{\Bigl(\prod\limits_i\DR_i\Bigr)}}
\newcommand{\intV}{\int_V}
\newcommand{\dirac}[1]{\ensuremath{\delta\bigl(#1\bigr)}}

\section{Monte Carlo algorithm\label{sec:method}}

We give a basic description of the algorithm previously developed in
Refs.~\cite{PRL88,JCP117,JCP120}. We first recall its
theoretical basis, and then present the simplest implementation,
highlighting places where the method can be further optimized for
speed.

\subsection{Theoretical foundation}

In order to sample configurations of a system containing charges, we
require that a set $\{\R_i\}$ of particle positions is generated with
weight
\begin{equation*}
  z\bigl(\{\R_i\}\bigr)
  =e^{-\beta\left[{\textstyle\frac{1}{2}\intV}
      \rho(\R;\{\R_i\})\Phi(\R;\{\R_i\})\DR
      \;+\;\mathcal{U}(\{\R_i\})\right]}
\end{equation*}
where $\rho\bigl(\R;\{\R_i\}\bigr)=\sum_iq_i\dirac{\R-\R_i}$ is the charge
distribution of the configuration, $\Phi\bigl(\R;\{\R_i\}\bigr)$ is the
unique solution to Poisson's equation, and $\mathcal{U}\bigl(\{\R_i\}\bigr)$
is the potential of all other interactions. The partition function then is
\begin{equation*}
  \mathcal{Z}=\int\DRi z\bigl(\{\R_i\}\bigr).
\end{equation*}

In the usual treatment of electrostatic interactions the electric
field is given by $\E_P=-\grad\Phi$: it is unique and satisfies both
Gauss's law $\div\E_P=\rho/\eps$ and the static version of Faraday's law
$\rot\E_P=\vect{0}$. We chose the convention where the potential of a
charge $q$ is $q/4\pi\eps r$. The algorithm is based on relaxing
Faraday's law so that $\E=\E_P+\rot\Q$, a decomposition familiar from
the Coulomb gauge of electrodynamics. Fourier transforming we find
that $\E_P$ is longitudinal:
\begin{equation*}
  (\E_P)_\k=-i\k\Phi_\k\parallel\k
\end{equation*}
and that $\rot\Q$ is transverse:
\begin{equation*}
  (\rot\Q)_\k=i\k\cross\Q_\k\perp\k.
\end{equation*}
As a consequence, the electrostatic energy
\begin{equation*}
  \begin{split}
    \frac{\eps}{2}\intV\E^2\DR
    &= \frac{\eps}{2}\left(\intV{\E_P}^2\DR+\intV(\rot\Q)^2\DR\right)\\
    &= \frac{1}{2}\intV\rho\Phi\DR+\frac{\eps}{2}\intV(\rot\Q)^2\DR
  \end{split}
\end{equation*}
so that the statistical weight of a configuration of charges and
field is
\begin{equation*}
  \begin{split}
    z'\bigl(\{\R_i\},\Etr\bigr)
    &=e^{-\beta\left[\mathcal{U}(\{\R_i\})
        +{\textstyle\frac{\eps}{2}\intV}\E^2\DR\right]}\\
    &=z\bigl(\{\R_i\}\bigr)
    e^{-\beta{\textstyle\frac{\eps}{2}\intV}\Etr^2\DR}
  \end{split}
\end{equation*}
where for clarity we have introduced the transverse field
$\Etr=\rot\Q$.  This field is constrained by $\div\Etr=0$; it is
independent of the charge configuration. Thus, the partition function
of the system of charges and field splits into two parts:
\begin{equation}
  \begin{split}
    \mathcal{Z}'
    &= \int\DRi\mathcal{D}\Etr\dirac{\div\Etr}z'\bigl(\{\R_i\},\Etr\bigr)\\
    &= \left(\int\DRi z\bigl(\{\R_i\}\bigr)\right)\\
    &\quad\times\left(\int\mathcal{D}\Etr\dirac{\div\Etr}
    e^{-\beta{\textstyle\frac{\eps}{2}\intV}\Etr^2\DR}\right)\\
    &= \mathcal{Z}\times\mathcal{Z}_\text{tr}
  \end{split}\label{eq:Z}
\end{equation}
where $\mathcal{Z}_\text{tr}$ is the partition function of transverse
field. The statistical weight of a configuration of charges is
$z\bigl(\{\R_i\}\bigr)\times\mathcal{Z}_\text{tr}$; all the weights have
been multiplied by the same constant. Hence configurational
probabilities are left unchanged. Of course sampling this system
requires introducing Monte Carlo moves appropriate for integrating
over $\Etr$ degrees of freedom.

\subsection{A charged lattice gas\label{sec:implementation}}

\begin{figure}[htb]
\includegraphics[width=\colwid]{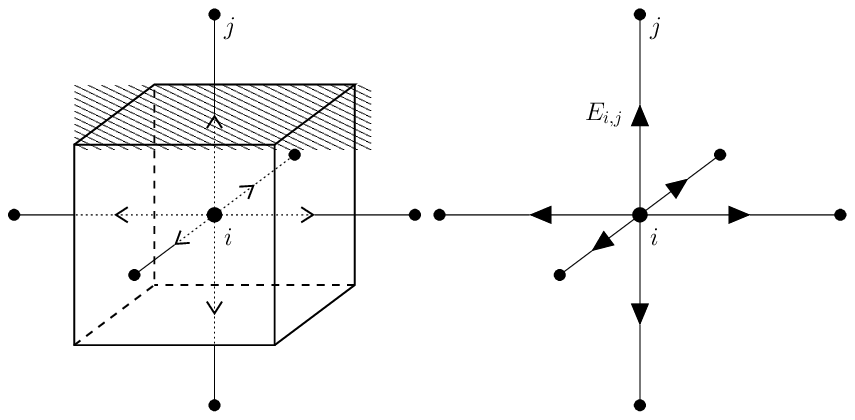}
\caption{\label{fig:node} Left: cubic lattice mesh around site $i$.  The
  electric flux $\phi_{i,j}$ flowing upward through the hashed cube face
  is assigned to link $(i,j)$ on the right: $E_{i,j}=\phi_{i,j}/a^2$.
  Field divergence at $i$ equals the outward flux through the cube
  surface, $\sum_{j\in\{\text{NN}\}}E_{i,j}$ where NN stands for nearest
  neighbors. In following figures only two dimensions of the lattice will
  be shown for clarity.}
\end{figure}

We consider a cubic simulation cell of $L^3$ sites with periodic
boundary conditions. Particles are placed on sites of a lattice with
mesh spacing $a$, and field
variables representing electric flux are defined on links. The
electric field divergence at a site is the sum of fluxes over the six
outgoing links.  (See Fig.~\ref{fig:node}.)  Transverse field
degrees of freedom appear as a nonzero line integral of $\E$ on
plaquettes of the lattice.

To start a simulation we must construct a state consistent with Gauss's
law. We initialize the electric field for the simulation with a single
sweep through the network: We use a procedure that follows a
\emph{Hamiltonian path} through the lattice.  Such a path visits each
site just once and traverses each link either once or zero times.  We
begin by initializing all field values on the lattice to zero and
start at an arbitrary point $1$ of the lattice; the node $1$ holds the
charge $q_1$. A single link of the path, $\{1,2\}$, connects it to
site $2$, on which we set the outgoing field to $q_1/a^2\eps$; Gauss's law
is now fulfilled on site $1$ and we move to the node $2$.

At each step, on arriving at site $i$ holding $q_i$, we have already
solved the Gauss constraint for sites $\{1,\hdots,i-1\}$. The incoming
link to the site, $\{i-1,i\}$, thus bears the initialized field
$E_{i-1,i}=\left(\sum_{j=1}^{j=i-1}q_j\right)/a^2\eps$. We now set the
outgoing field $E_{i,i+1}$ to $\left(\sum_{j=1}^{j=i}q_j\right)/a^2\eps$
so that $E_{i,i-1}+E_{i,i+1}=q_i/a^2\eps$: Gauss's law is now fulfilled on
site $i$ and we go to site $i+1$. At the end of the path, we reach
site $V=L^3$ with $E_{V-1,V}=\left(\sum_{j=1}^{j=V-1}q_j\right)/a^2\eps$.
The imposition of periodic boundary conditions in charged systems is
only possible if the total charge $Q$ is zero (otherwise the total
energy is divergent).  Thus $E_{V-1,V}=(Q-q_V)/a^2\eps=-q_V/a^2\eps$ and
Gauss's law is satisfied everywhere on the lattice.

\begin{figure}[htb]
  \includegraphics[width=\colwid]{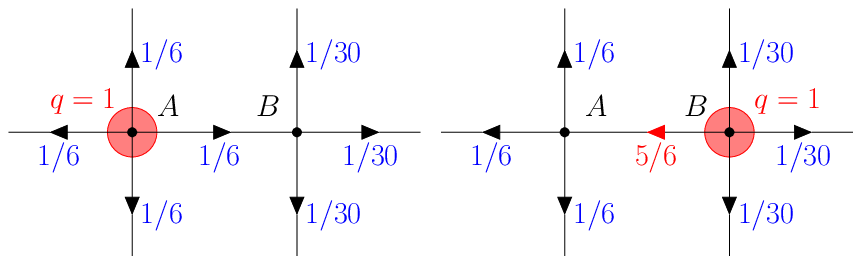}
\caption{\label{fig:move}
  A pair of lattice sites, before and after a particle move.  Left, the
  initial configuration is made up of a charge at $A$ and \emph{one}
  solution to Gauss's constraint: at $A$ the field divergence is six
  times $1/6$, equaling $q=1$ (in reduced units where $a=1$, $\eps=1$),
  and at $B$ it is $5\times1/30-1/6=0$.  Right, the charge has moved to
  $B$ and a flux $\delta E=q=1$ flowing from $B$ to $A$ has been added
  to the central link.  Then Gauss's law is again verified: at $A$,
  $\div\E=5\times1/6-5/6=0$, and at $B$, $\div\E=5\times1/30+5/6=1=q$.}
\end{figure}

\newcommand{\rhoi}{\rho^{(\text{i})}}
\newcommand{\rhof}{\rho^{(\text{f})}}
\newcommand{\Ei}{\E^{(\text{i})}}
\newcommand{\Ef}{\E^{(\text{f})}}
We take advantage of the new field degrees of freedom to construct
local updates for charge moves.  Consider an initial configuration
$\rhoi$ where a charge $q$ is at point $A$, and the initial electric
field $\Ei$ satisfies Gauss's law ($\div\Ei=\rhoi/\eps$). If a trial
places $q$ at point $B$, the final configuration is
$\rhof=\rhoi-q\dirac{\R-\R_A}+q\dirac{\R-\R_B}$, and a new solution
for the field must be found. In order to remain consistent with Gauss's
law it is sufficient to add to $\Ei$ field lines $\delta\E$ flowing
from $B$ to $A$ and totalizing $q/\eps$ flux.  The final field
$\Ef=\Ei+\delta\E$ now satisfies again Gauss's law for the final
charge configuration:
\begin{equation*}
\div\Ef =\frac{\rhoi}{\eps}
+\frac{q}{\eps}\dirac{\R-\R_B}-\frac{q}{\eps}\dirac{\R-\R_A}
=\frac{\rhof}{\eps}.
\end{equation*}
We call $\delta\E$ ``the slaved update''. Nothing having been required
of it except the total flux, we can choose it to be localized in
space, so that charge moves result in local updates of the simulated
system.  In addition $\delta\E$ should be symmetrically chosen so that
detailed balance applies to the forward and reverse updates.  Our
previous choice of $\delta\E$, which modifies just the link connecting
$A$ to $B$ is illustrated in Fig.~\ref{fig:move}.  Nevertheless, the
``total flux'' constraint lets us free to use more complex slaved
updates. We will show in this paper that splitting $\delta\E$ into
several lines helps increase the algorithm efficiency.

\newlength{\mylen} \setlength{\mylen}{\colwid /2 - \tabcolsep}

\begin{figure}[htb]
\includegraphics[width=\colwid]{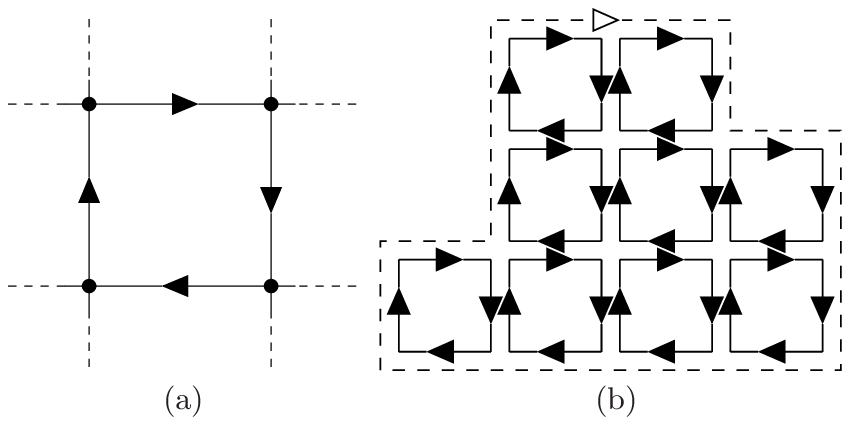}
\caption{\label{fig:plaq} To update $\Etr$ one may choose between
  plaquette moves (a) which increase the field along a single plaquette
  edge, and worm moves [(b), dashed line] which modify it along the path
  of a given random walk. Both are thermodynamically equivalent since
  any field configuration reached through worm moves may be obtained
  through multiple plaquette updates [as shown in (b)] combined with
  updates of the $k=0$ mode of the field.
}
\end{figure}

Finally in order to correctly sample the partition function
\eqref{eq:Z} we integrate over the transverse degrees of freedom of
the electric field.  We do this with Monte Carlo moves which change
the circulation of the electric field, but do not modify its
divergence.  One way of doing this is by modifying the field on the
four links defining a plaquette [Fig.~\ref{fig:plaq}(a)]. If one
increases the field on links along the edge of a given plaquette by
some constant value, at all sites $\div\E$ remains constant.
This kind of update, being local, leads to diffusive dynamics for
$\Etr$: $\ord{L^2}$ sweeps are needed to yield an independent
configuration.


An alternative method of integrating over $\Etr$ was introduced
\cite{StatPhys}. These \emph{worm} updates [Fig.~\ref{fig:plaq}(b)]
make use of a biased random walk to generate a closed contour along
which the field is modified. This contour visits typically $L^3$ sites
and turns out to be particularly efficient at equilibrating the
electric field at all length scales simultaneously:
all Fourier modes of $\Etr$ decay at the same rate, $\ord{1}$ sweeps
are enough to produce an independent field configuration.

\newlength{\mylenb} \setlength{\mylen}{\colwid * \real{0.425}}
\setlength{\mylenb}{\colwid - \tabcolsep * 2 - \mylen}

The aim of a Monte Carlo algorithm is to produce statistically
independent configurations with minimum computational cost. The local
updates described above allow one to efficiently update charge and
field configurations. However in order to understand the global
dynamics and convergence of the algorithm we shall study
electric field autocorrelation functions.
We now show that high mobility $\mu$ of the charges leads to fast
decay of the field correlations.

\newcommand{\dt}{\ensuremath{\delta t}}

\section{Measuring charge mobility\label{sec:mobility}}


Under the dynamics of the algorithm, Figs.~\ref{fig:move} and
\ref{fig:plaq}, $\E$ remains consistent with Gauss's law at all times.
Considering the time derivative of this law, we find
\begin{equation*}
  \div\frac{\partial\E}{\partial t}=\frac{\partial(\div\E)}{\partial t}
  =\frac{1}{\eps}\frac{\partial\rho}{\partial t}
\end{equation*}
which translates in the implementation as
\begin{equation}
  \div\delta\E=\frac{1}{\eps}\delta\rho.\label{eq:timegauss}
\end{equation}
Updates to the electric field can be considered as being due to local
currents such that
\begin{equation}
  \div\J + \frac{\delta\rho}{\dt} =0, \label{eq:charge}
\end{equation}
where we introduced the time unit $\dt=1$~Monte Carlo step.
Combining Eqs.~\eqref{eq:timegauss} and \eqref{eq:charge} we find
\begin{equation*}
  \div\delta\E=-\frac{\dt}{\eps}\,\div\J,
\end{equation*}
or
\begin{equation}
  \delta\E=-\frac{\dt}{\eps}\J+\dt\,\rot\vect{H}  \label{eq:evol}
\end{equation}
with $\vect{H}$ arbitrary. This is a discrete version of Ampere's law
of electromagnetism.

Spatially averaging Eq.~\eqref{eq:evol}, we find the change in the average
electric field $\bar{\E}$ during an update,
\begin{equation}
  \delta\bar{\E}=-\frac{\dt}{\eps}\bar{\J}. \label{eq:ebar}
\end{equation}
This equation is independent of $\vect{H}$; the last term in
Eq.~\eqref{eq:evol} gives zero due to periodic boundary conditions.  This
is consistent with the fact that local plaquette updates do not change
the average electric field in a periodic system.

Our simulations are on a system containing $N$ mobile unit charges, either
the symmetric plasma made up with $N/2$ particles of each sign ($q_i
=±e$), or the one-component plasma (OCP) of $N$ positive charges
moving in a fixed negative background.  Linear response gives insight
on the relation between charge mobility and field evolution. The
electric current is due to the movement of mobile charges,
\begin{equation}
  \J=\sum_i\J_i=\sum_i\rho_i\vect{v}_i=\sum_i q_i n_i\vect{v}_i, \label{eq:J}
\end{equation}
where $i\in\{+,-\}$, $\rho_i$ are charge densities and $n_i$ are
number densities; $n_- =0$ for the OCP. On average, velocities are
related to field by
\begin{equation}
  \vect{v}_i=\mu q_i\E. \label{eq:mu}
\end{equation}
Given the charge symmetry of the algorithm, positive and negative ions
in a symmetric plasma have the same mobility. Equations~\eqref{eq:J} and
\eqref{eq:mu} lead to $\J=e^2(n_+ + n_-)\mu\E$.  $n_+ + n_-=n=N/V$ is
the number density of mobile charges. Hence
\begin{equation}
  \J=e^2n\mu\E. \label{eq:JE}
\end{equation}
We should bear in mind that these relations are phenomenological. For
example, in Eq.~\eqref{eq:mu} proportionality holds only when the field
intensity is not too high. It will also become apparent that in
certain limits $\mu$ can fall to zero for large, dilute systems.

Substituting Eq.~\eqref{eq:JE} in Eq.~\eqref{eq:ebar}, and replacing the
difference equation by a differential equation we find that
\begin{equation}
  \frac{\partial\bar{\E}}{\partial t}=
  -\frac{e^2n\mu}{\eps}\bar{\E}.\label{eq:diffebar}
\end{equation}
$\bar{\E}$ is the $\k=\vect{0}$ Fourier mode of the electric field.
Equation~\eqref{eq:diffebar} implies that the autocorrelation function of this
mode behaves as follows:
\begin{equation*}
  \left\langle\bar{\E}(t')\bar{\E}(t'+t)\right\rangle_{t'}
  =Ce^{-\tfrac{e^2n\mu}{\eps}t},
\end{equation*}
where $C$ is the squared amplitude of the thermal fluctuations of
$\bar{\E}$.  Measuring this autocorrelation function we find
exponential decay with a characteristic time $\tau_0=\eps/(e^2n\mu)$
or equivalently a decay rate $\lambda_0=e^2n\mu/\eps$. We fit all our
numerical data with a single exponential and verify the quality of the
resulting curve by eye.

In our simulations we also monitored other modes of the field and
found that the mode $\k=\vect{0}$ is the slowest. Higher modes of the field
couple directly to plaquette updates as well as particle motion, and
relax with the dispersion law $\lambda_k=\lambda_0+D_Ek^2$
\cite{JCP117}. Larger $k$ are less sensitive to low particle mobility
(low $\lambda_0$).  They will not be considered further in this
paper.

The time scale $\tau_0$ can be understood with a scaling argument. In
order to produce two uncorrelated samples of the system, one should
wait for the charges to diffuse through the characteristic correlation
length of the system, the Debye length $l_D=\sqrt{\eps k_BT/e^2n}$.
Thus $\tau_0={l_D}^2/D$, and $\lambda_0={\kappa_D}^2D=e^2nD/\eps
k_BT$, with a diffusion constant $D$.  We recover the above expression
for the relaxation rate if we use the relation $D=k_BT\mu$.  $D$
defined in this way relates to the mobility of charges under an
external electric field (where opposite charges move opposite ways),
\emph{not} to the mobility under a constant external force (where all
particles move together).

Thus we will measure the mobility of charges or, equivalently, their
diffusion coefficient, by computing the autocorrelation function
of the average electric field. This method has the additional advantage
that we need not keep track of the winding of particles across the
periodic cell boundaries.

\section{Limiting factors for mobility\label{sec:interpret}}

Acceptance rate is often used to monitor the efficiency of Monte Carlo
simulations.  However a high acceptance rate does not necessarily mean
useful work has been performed.  For example, the diffusion dynamics
of point defects or interstitials in a crystal are very slow. In a
Monte Carlo simulation most time is spent vibrating the atoms around
their equilibrium positions; even in the limit of very rare diffusion
events the acceptance rate of trial moves remains appreciable.
Another example is magnetization reversal of an Ising ferromagnet. The
state where all spins are oriented against an applied field is
metastable but with very long lifetime; since the Metropolis algorithm
is already very inefficient one uses rejection-free algorithms
\cite{novotny} to update individual spins, unfortunately after a spin
has been flipped it is almost certainly flipped back at the next step,
so that the magnetization never reverses within accessible simulation
times.

\begin{figure}[hbt]
  \includegraphics[bb=0 0 226 163,clip=false,width=\colwid]{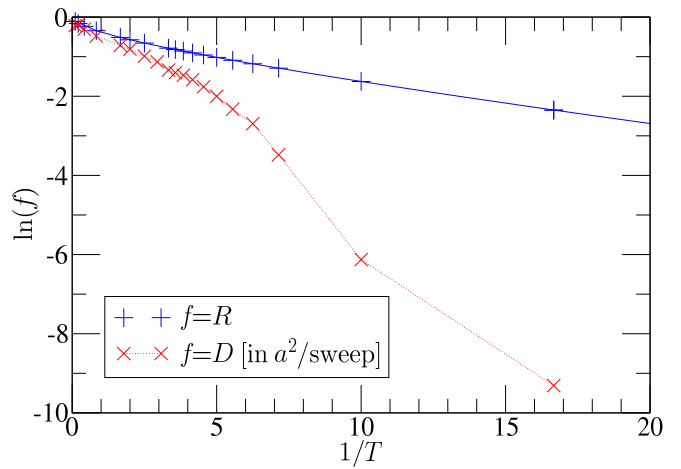}
  \caption{\label{fig:acc+gap}
    Logarithm of acceptance rate $R$, and of diffusion coefficient $D$
    expressed in $a^2$ per particle sweep, versus inverse temperature.
    Solid line, Eq.~\ref{eq:R}; dotted line, guide to the eye.
    $R$ and $D$ are close to $1$ at high temperatures, but $D$ drops
    much faster than $R$ on decreasing $T$. One component plasma of
    two positive unit charges, box of size $L=15$.}
\end{figure}

With our algorithm, a simulation performed at very low density,
Fig.~\ref{fig:acc+gap},
shows that the diffusion coefficient $D$ of
charges drops much faster at low temperatures than the acceptance rate
of particle moves: atoms simply wander around their mean locations,
rather like in the examples above.

\subsection{Variation of the acceptance rate\label{sec:ptlacc}}

In the algorithm summarized in Sec.~\ref{sec:implementation},
motion of a charge modifies the field on the single link along which
the particle has moved (see Fig.~\ref{fig:move}).  Let $\Elk$ be the
field intensity on this link before the move. The divergence at $A$ is
$q/a^2\eps$, and since in the absence of other nearby charges $\E$ must be
isotropic around $A$ we expect that $\langle\Elk\rangle =q/6a^2\eps$.
Fluctuations of $\Etr$ imply that $\Elk=q/6a^2\eps+\eta$, with $\eta$ a
Gaussian random variable with standard deviation $\sigma$. The energy
in these fluctuations is $3L^3a^3\left\langle\frac{\eps}{2}\eta^2\right\rangle
=\frac{3}{2}\eps L^3a^3\sigma^2$.  From
equipartition and given that there are two polarizations of $\Etr$,
the energy in $\Etr$ is also approximately $L^3k_BT$, thus we conclude
that $\sigma^2=2k_BT/3a^3\eps$.
\begin{figure}[t]
  \includegraphics[bb=0 0 226 163,clip=false,width=\colwid]{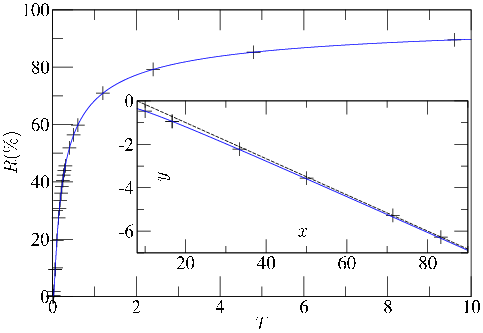}
  \caption{\label{fig:ln_acc}
    Acceptance rate $R$ of charge moves versus temperature. +, simulation
    results; solid line, Eq.~\ref{eq:R}. Inset, $y=\ln(R/\sqrt{T})$
    against $x=1/T$.  Numerical results approach the asymptote
    Eq.~\eqref{eq:asymp} of slope $-1/12$ (dashed line).  A pair of opposite
    charges, box size $L=15$.}
\end{figure}

During motion of the charge, the field on $AB$ is modified to
$-5q/6a^2\eps+\eta$. The energy difference between the two configurations
is thus $\delta\mathcal{E}=qa(q/3a^2\eps-\eta)$.  With the Metropolis
algorithm when $\eta<q/3a^2\eps$ the trial is accepted with probability
$\exp(-\delta\mathcal{E}/k_BT)$, otherwise it is automatically
accepted. Computing the average over all values of $\eta$, we find the
acceptance rate
\begin{equation}
R=\mathrm{erfc}\left(q/2\sqrt{3a\eps k_BT}\right). \label{eq:R}
\end{equation}
We plot this function together with numerical results in
Fig.~\ref{fig:ln_acc}.  When $T$ is small, the asymptotic expansion
of $\mathrm{erfc}$ gives
\begin{equation}
R=\sqrt{\frac{12T}{\pi}}e^{-1/12T}\bigl(1+\ord{T}\bigr). \label{eq:asymp}
\end{equation}
Defining $y=\ln\bigl(R/\sqrt{T}\bigr)$ and $x=1/T$, we find an Arrhenius
law for the acceptance rate $y=-x/12+\text{const.}+\ord{1/x}$, which is
illustrated in the inset of Fig.~\ref{fig:ln_acc}.

We conclude that particle motion becomes hard with this update scheme
for $T<1/12$ due to a finite energy barrier.  One of our aims in the
rest of this paper will be to reduce the barrier so that the
acceptance rate remains high even for temperatures $T\ll 1/12$.

\subsection{Field trails and string tension\label{sec:trails}}

\begin{figure}[b]
  \includegraphics[width=\colwid]{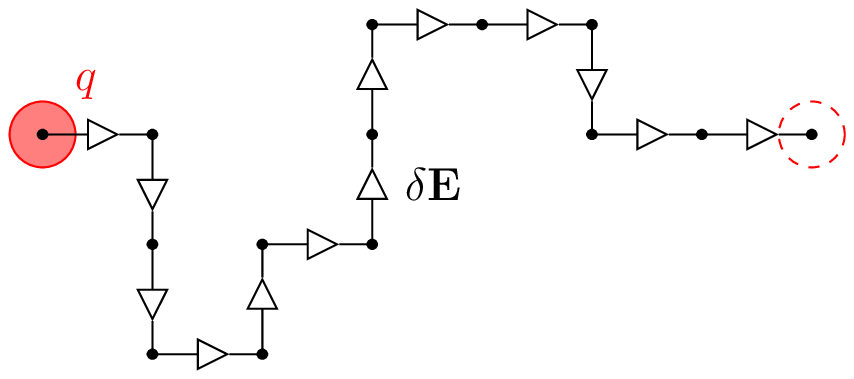}
  \caption{\label{fig:string}
    The field update produced by successive moves of a charge $q$ is a
    field string of intensity $\delta E=q/a^2\eps$ connecting the
    particle to its starting position (dashed circle). The trail
    remains as long as no plaquette update intervenes.}
\end{figure}

Let us consider two closely separated charges with the electric field
in equilibrium. The field has the usual dipolar form familiar from
elementary electrostatics. What happens if we pull very hard on the
positive charge so that the separation between the charges increases
rapidly, without updating the plaquette degrees of freedom?  During
the motion of the particle each link traversed is modified by
$q/a^2\eps$ leaving behind a ``trail'' of modified links
(Fig.~\ref{fig:string}).
With time the field configuration will
relax back to a dipolar form because of the updates of the plaquettes
equilibrating the transverse field.  However on a short time scale
there are few plaquette updates, and dragging the charge along $r$
links costs an energy which we can estimate to be $r\gamma_0$, where
\begin{equation}
  \gamma_0= q^2/2a\eps \label{eq:tension}
\end{equation}
is our estimate of the energy rise per link,
\begin{equation*}
  a^3\frac{\eps}{2}\left[\left(E-\frac{q}{a^2\eps}\right)^2-E^2\right]=-aqE+\frac{q^2}{2a\eps}
\end{equation*}
and $E$ has
zero mean.  There is a ``string tension'', $\gamma_0$, pulling
the particles back.

In the presence of an external electric field, a pair of opposite
charges normally separates. A finite string tension implies that
this mobility is suppressed. One must spend much numerical effort on
updating the plaquettes in order to destroy the trail and stop
particles backtracking.  While the string tension is positive at low
temperatures we will now argue that the thermodynamic tension $\gamma$
should become zero at a finite temperature: above it the
mobility is high even at low frequencies of updates in the plaquettes.

Consider a trail joining two fixed test charges separated by a
distance $r$ and let the length of the trail joining them be $\ell$.
If $\ell\gg r$ we can estimate the number of such paths from
the statistics of the path: $N_\ell=\ord{z^\ell}$. $z$ is a connectivity
constant characterizing the geometry of the walk. For a random walk
$z=6$, for a self-avoiding walk $z=4.68$.  We now estimate the free
energy of the configuration as
\begin{equation}
F\approx \ell\gamma_0 - k_BT\ell\ln z = \gamma\ell.\label{eq:peierls}
\end{equation}
From this expression we can expect two distinct dynamic regimes for
the algorithm. At low temperatures the tension $\gamma$ is positive
and it is most favorable for the trail to remain short, $\ell \sim r$.
The free energy for separating the charges is indeed linear and we
have a phenomenon similar to confinement in gauge theories. This
confinement is only destroyed by the dynamics of the plaquettes which
slowly relaxes the trail into a dipolar field configuration.  At
temperatures higher than $T_c\sim\gamma_0/k_B\ln z\approx 0.3$ the
line tension drops to zero and the particles become unconfined.  Even
without plaquette updates the particles remain mobile and can separate
easily.

We also note that there is a very close analogy between this picture
of roughening trails and the ($2+1$)-dimensional Hubbard model in the
phase approximation, which can be expressed as a set of fluxes on a
lattice \cite{young,alet}. This model has two thermodynamic phases,
one with tense field lines which are strongly suppressed, and a
superconducting phase in which field lines proliferate. The transition
occurs at a temperature $T\approx 0.33$.

In Fig.~\ref{fig:acc+gap} we used a split in which one half of all
updates try to move one of the two particles, and one half of updates
modify a randomly chosen plaquette. The number of plaquettes
($3L^3\approx 10^4$) is much larger than the number of particles
($N=2$), so that a given plaquette is rarely updated, trail
formation is probable.  The diffusion coefficient of charges indeed
drops at a crossover temperature $T_c\approx 0.2$ which
qualitatively agrees with the above estimate.

\begin{figure}[hbt]
  \includegraphics[width=\colwid]{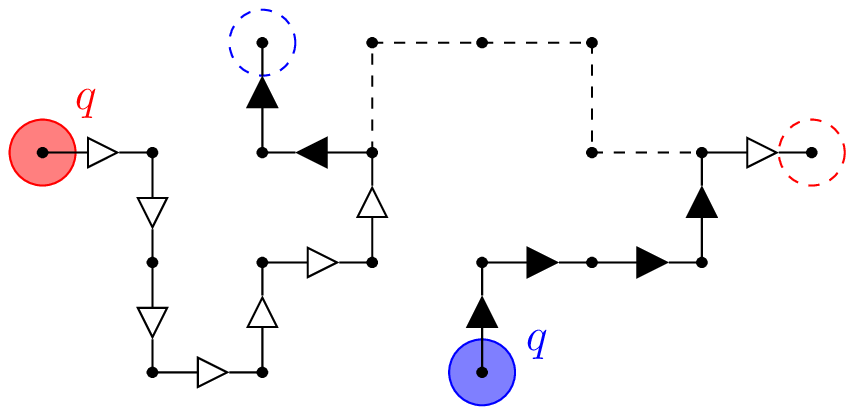}
  \caption{\label{fig:caught}
    A second charge joins the string left by the charge of
    Fig.~\ref{fig:string}; it is dragged along the original path.
    Field updates then erase the previous trail (dashed line).  The
    field strings no longer connect the particles to their respective
    starting sites (dashed circles).  }
\end{figure}

How do we expect this trail-limited mobility to vary as a function of
charge density?  If a charge $i$ creates a trail, and a charge $j$ of
the same sign crosses it, then $j$ will also feel the mean force
mentioned above.  If $j$ is now dragged back along the track of $i$,
the field updates will erase the trail (Fig.~\ref{fig:caught}).
Afterwards, neither $i$ nor $j$ are linked to their initial positions.
We thus expect that the effect of the trails is cut off at a distance
comparable to the inter-particle spacing.  Indeed we do find that the
mobility increases on simulating systems of increasing charge
densities.  Thus in this paper we will concentrate on improving the
efficiency of the algorithm at very low densities, working most often
with samples containing just two charges.

In the next two sections we will modify the slaved updates in order
to reduce the bare tension of strings. With a lower $\gamma_0$ we will
lessen the crossover temperature $T_c\approx 0.2$ which results from
the balance between energy and entropy expressed in Eq.~\eqref{eq:peierls}.
To efficiently simulate condensed matter systems, particles must remain
highly mobile to much lower temperatures: $T=0.2$
corresponds to $l_B=a/4\pi T\approx 0.4~\text{Å}$ (with $a=1~\text{Å}$),
whereas the Bjerrum length in water at room temperature is $l_B
\approx 7~\text{Å}$.  Therefore we aim at lowering $T_c$ by a factor
of approximately $20$.

\section{Extended charges\label{sec:spr}}

The expression \eqref{eq:tension} for the bare tension of the string
is \emph{quadratic} in the charge, $\gamma_0=q^2/2a\eps$.  Let us now
split the string between two particles into $K$ substrings; each
substring carries a flux of $q/K\eps$.  The bare tension of each
substring will be $\gamma_0/K^2$, and that of the whole split string
will be $K(\gamma_0/K^2)=\gamma_0/K$.

\begin{figure}[hbt]
  \includegraphics[bb=0 0 226 163,clip=true,width=\colwid]{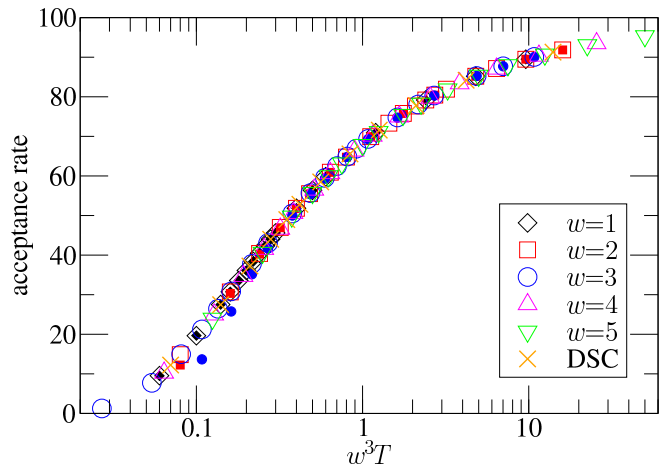}
  \caption{\label{fig:acc}
    Acceptance rate of particle moves versus temperature for OCP (open
    symbols, $w=1$ to $5$) and a pair of charges (filled symbols,
    $w=1$ to $3$), temperature rescaled by $w^3$ (DSC, $7$).}
\end{figure}

In this section, to form split strings we spread the particles on cubes
of side $w$; each site in the cube carries a subcharge of $q/w^3$, and
when a particle moves the field is updated on the $w^3$ links crossed by
each subcharge. We use values
of $w$ ranging from 1 (the original algorithm) to 5, and measure the
acceptance rate of particle updates. When we plot the rate as a
function of $w^3T$ (Fig.~\ref{fig:acc}) we find that all curves
collapse, except at low temperatures for the two opposite charges due
to pairing.  We also simulated point charges with the coupled update
proposed by Duncan, Sedgewick, and Coalson in Ref.~\cite{coalson} (hereafter
denoted by ``DSC''), and the acceptance rates collapse equally well.

The scaling of the acceptance rate in $w^3$ can be understood as follows:
motion of each part of the particle is hindered by a barrier which varies as
$(1/w^3)^2$. The barriers are additive leading to a local barrier with an
amplitude which varies as $1/w^3$.

\begin{figure}[hbt]
  \includegraphics[bb=0 0 226 163,clip=true,width=\colwid]{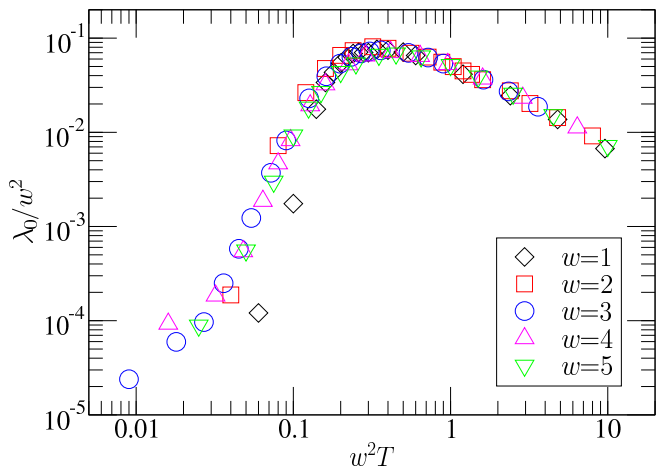}
  \caption{\label{fig:spr}
    Mobility in OCP versus temperature ($N=2$, $L=15$). Each charge is
    spread on a $w$-site-side cube.  Data collapse when temperature and
    $\lambda_0$ are
    scaled by $w^2$.  For $T\rightarrow\infty$ the diffusion
    coefficient of particles is bound by $1~a^2\,(\text{sweep})^{-1}$:
    $D$ saturates and $\lambda_0\sim 1/T$.}
\end{figure}

When we plot mobility (determined from the dynamics of $\bar{\E}$) as
a function of temperature, we find that the benefit obtained from
charge spreading is \emph{not} proportional to $w^3$; curves collapse
on using a scaling with $w^2$, Fig.~\ref{fig:spr}.  The cross-section
area of the extended charges is equal to $w^2$, so their field trails
are made up from $K=w^2$ field lines of strength $q/a^2w^2\eps$. This
gives a bare tension for the trail of $\gamma_0/w^2$.  When trail
formation limits mobility, the typical crossover temperature $T_c$
thus scales as $1/w^2$. This seems to indicate that the statistics
of the paths and the connectivity constant do not change with $w$.

To further confirm the idea that field trails are
limiting mobility we introduced a new kind of field update:
We define a cubic box of side $b$ centered on a site occupied by a
particle, and then generate a \emph{worm} update
(Sec.~\ref{sec:implementation}, and Ref.~\cite{StatPhys}) inscribed
in the box.  At each Monte Carlo step, the algorithm attempts one of
three updates, either a particle move, a plaquette update, or a
``local worm''.  Since all choices are reversible detailed
balance is verified.

Worm moves are known to lead to fast relaxation of the field, so these
new ``local worm updates'' should allow one to spread out the field
trails efficiently, concentrating the computational effort around
charges, where the trails are formed. By introducing them in a 1:1
proportion with particle moves (with $b$ satisfying $b^2>w^3$), we
expect to cancel the effective string tension.
This computation is very expensive; one ``local worm update'' is far
more costly than one particle update. We did not seek further
optimization, and do not recommend this method for production of data
with the algorithm.

\begin{figure}[hbt]
  \includegraphics[bb=0 0 226 163,clip=true,width=\colwid]{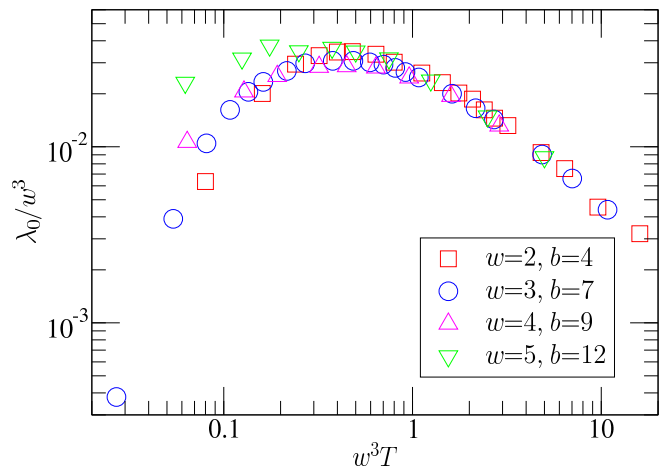}
  \caption{\label{fig:lw}
    Mobility of OCP particles versus temperature at low density ($N=2$,
    $L=15$). Here local worm moves are introduced (see text), and now data are
    made collapse when scaled by $w^3$, showing that field trails have been
    removed, as opposed to data of Fig.~\ref{fig:spr}.}
\end{figure}

We find that the crossover
temperature $T_c$ of the mobility drop decreases with these local worm
moves.  In Fig.~\ref{fig:lw} the data superimpose if we rescale by a
factor of $w^3$, implying that the trails are no longer dominating the
dynamics.
The $w^3$ scaling may be indicating that dynamics are now limited by
the local barrier to particle hops described in Sec.~\ref{sec:ptlacc}.


The spreading of particles over several sites clearly modifies the
interactions at short distance. One should introduce a hard core
interaction for distances less than $wa$, corresponding to the
diameter $r_0$ of the particles.  Much of the interesting physics in
soft condensed matter depends on the \emph{ratio} of the Bjerrum
length $l_B=e^2/4\pi\eps k_BT$ to the particle size. One is typically
interested in the range $5<l_B/r_0<20$, which corresponds to
$0.004<wT<0.02$. While we have succeeded in reducing the crossover
temperature $T_c$ by a factor $1/w^2$ we have also changed the
physical length scale by a factor $w$.  The final result is only a
factor $w$ improvement in $T_c$ when measured in physical units;
a lattice algorithm suitable for condensed matter
simulation would require $w\approx 20$.  Such fine discretization has
been used in lattice models to reproduce correctly thermodynamical
properties of some systems \cite{panawater,panamacro}. However for
cases where this is not required, one might prefer to avoid such
large $w$.
We now explore methods of moving charges which do
not require permanent spreading so that the effective length scale in
the simulation is not modified.

\section{Temporary charge spreading\label{sec:flum}}

\begin{figure}[hbt]
  \includegraphics[width=\colwid]{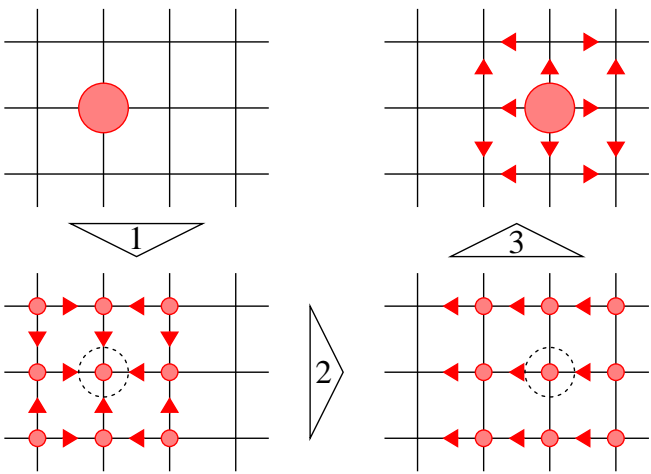}
  \caption{\label{fig:floumpf}
    Temporary spreading of charge (here with $w=3$).  First step,
    spread the charge to $w^3$ sites. Second step, move the extended
    set of charges.  Third step, inverse of the first; charge
    fractions collapse back to a point. The overall field update is
    the sum of the individual steps.}
\end{figure}

There is a direct way of reconciling the requirements that charges are
extended during their motion but otherwise pointlike: Before moving a
particle, one should first spread its charge evenly onto neighboring
sites, then move all subcharges as a block, and finally bring them
back together (see Fig.~\ref{fig:floumpf}). This defines a
charge move involving three substeps.

Each step consists of a set of currents. When a charge is split a
current $\j^{(1)}$ flows from the central site. Motion of the particle
generates a current, $\j^{(2)}$. When the charge is collapsed to a
point a current $\j^{(3)}$ flows from the neighboring sites back to
the center. To maintain the constraint of Gauss's law, each of these
currents $\j^{(\alpha)}$ is associated with a field update
$\delta\E^{(\alpha)} =-\j^{(\alpha)}\dt/\eps$. For step 2, the current on
each modified link is $j^{(2)}=q/w^3a^2\dt$, as above. During step 1 the
values $j_i^{(1)}$ of the current on links $\{i\}$ are
under-determined, they are constrained only by charge conservation
Eq.~\eqref{eq:charge}. We thus additionally require that
$\sum_i\bigl(j_i^{(1)}\bigr)^2/2$ be a minimum, giving a unique, reversible
recipe for the current. We solve for $\j^{(1)}$ by minimizing the
functional
\begin{equation*}
  \mathcal{F}=\int\frac{\left(\j^{(1)}\right)^2}{2}
  -\Lambda\left(\div\j^{(1)}+\frac{\delta\rho^{(1)}}{\dt}\right).
\end{equation*}
The current is the solution of a Poisson-type problem,
\begin{equation*}
  \j^{(1)}=-\grad\Lambda\quad\text{and}\quad\nabla^2\Lambda
  =\frac{\delta\rho^{(1)}}{\dt}
\end{equation*}
with $\j^{(1)}\cdot\vect{n}=0$ on the boundary of the spread charge. The
solution to this equation is computed once during initialization of
the simulation and stored in a lookup table. Step 3 is the exact
reverse of step 1, $\j^{(3)}=-\j^{(1)}$.
On adding the fields $\j^{(1)}$, $\j^{(2)}$, and $\j^{(3)}$ we find a
flow going from the starting site to the final site, and taking
several paths.

\begin{figure}[hbt]
  \includegraphics[bb=0 0 226 163,clip=true,width=\colwid]{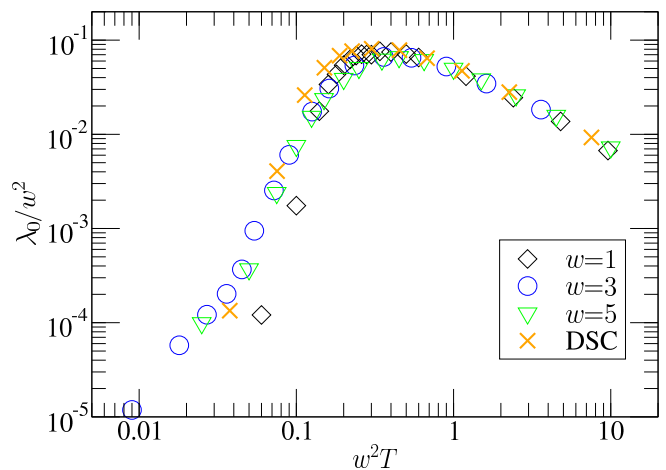}
  \caption{\label{fig:flum}
    Mobility of temporarily spread charges versus temperature.  OCP,
    $N=2$, $L=15$.  The mobility with no spreading ($w=1$) is also
    given. Data are rescaled by $w^2$ (DSC, $3.77$). }
\end{figure}

If now we simulate our test system using this version of the algorithm
and plot the mobility of particles versus temperature
(Fig.~\ref{fig:flum}), we find practically the same results as in
Fig.~\ref{fig:spr}. The crossover temperature scales with
$1/w^2$. The advantage is that the particles are still pointlike
unlike Sec.~\ref{sec:spr}. Thus we have improved the
lowest temperatures efficiently accessible by a factor of $w^2$ without
changing the physically important length scale.

The method has the advantage of both simplicity and generality.
\begin{itemize}
\item A small Poisson equation is solved once before starting simulation
  in order to determine the ``current map''.
\item One is free to choose as an intermediate state an arbitrary charge
  cloud.
\end{itemize}

We note that temporary spreading includes the Monte Carlo algorithm of
DSC as a special case, it is sufficient to consider the six nearest
neighbors of a site, plus the site itself, as the volume over which
the charge is to be spread.  Each site thus gets one seventh of the
total charge $q$, which explains the scaling of acceptance rate in
Fig.~\ref{fig:acc}. The result of steps 1 through 3 of
Fig.~\ref{fig:floumpf} yields a current $3q/7a^2\dt$ on the center link
and $q/7a^2\dt$ around the four plaquettes adjacent to it.  The curve
obtained by using DSC updates collapses with the others in
Fig.~\ref{fig:flum}, when scaled by $49/13=3.77$ which comes from a
simple estimate of the bare string tension.  However in our
simulations we have not implemented the additional step of simulating
the update with heat bath rather than Metropolis update.

\newcommand{\C}{\mathcal{C}}
\newcommand{\DE}{\Delta\mathcal{E}}
Rather than performing three successive steps, another implementation
of the temporary spreading consists in precomputing the total ``map''
of currents $\j^{(1)}+\j^{(2)}+\j^{(3)}$.  Such an implementation is
faster, avoiding multiple updates of the same links through steps 1--3.
However, keeping the three sub-steps distinct allows one to perform
several intermediate updates of step 2, moving the particle a
large distance before ``recondensing'' it to a point, as follows.
Consider a (starting) configuration $\C_\text{S}$ of the system. We
randomly choose one particle and spread its charge unconditionally;
the field is updated accordingly and we label the new configuration as
$\C_{i=0}$; the energy difference between $\C_\text{S}$ and $\C_0$ is
stored. Then we successively try $d$ moves of that particle in random
directions, which lead to configurations $\C_i$, $1\leqslant i
\leqslant d$. The field is updated and trials are accepted with
the Metropolis probability
\begin{equation*}
  m(\DE)=\min\left(1,\exp\left(-\frac{\DE}{k_BT}\right)\right)
\end{equation*}
where $\DE$ is the energy difference between the tried and current
configurations. Finally the charge is condensed, yielding the ending
state $\C_\text{E}$, and the whole update is accepted with probability
\begin{equation*}
  p_\text{acc}
  =m\bigl(\DE(\C_\text{S}\rightarrow\C_0)
  +\DE(\C_d\rightarrow\C_\text{E})\bigr).
\end{equation*}
The method saves computational effort, for each series of $d$ moves
there are only one spreading and one condensation steps; there would
be $d$ such steps if the procedure of local hopping of
Fig.~\ref{fig:floumpf} were used.

To prove that detailed balance is obtained, consider an instance of
such an update. Its global probability is
\begin{equation*}
  p(\C_\text{S}\rightarrow\C_\text{E})=p_\text{acc}
  \prod_{i=0}^{d-1}p(\C_i\rightarrow\C_{i+1}).
\end{equation*}
The probability of the ($i+1$)th step is
\begin{multline*}
p(\C_i\rightarrow\C_{i+1})=\\
  \begin{cases}
    \displaystyle\frac{1}{6}\,m\bigl(\DE(\C_i\rightarrow\C_{i+1})\bigr)
    & \text{if $\C_i\neq\C_{i+1}$ (accepted trial),}\\
    \displaystyle 1
    -\frac{1}{6}\sum_{\alpha'}m\bigl(\DE(\C_i\overset{\alpha'}{\rightarrow})\bigr)
    & \text{if $\C_i=\C_{i+1}$ (rejected trial),}
  \end{cases}
\end{multline*}
where the sum runs over the six directions of space, and
$\DE(\C_i\overset{\alpha'}{\rightarrow})$ is the energy change
corresponding to a trial move in the $\alpha'$ direction from
configuration $i$. The probability of the reverse update reads
\begin{equation*}
  p(\C_\text{E}\rightarrow\C_\text{S})=p'_\text{acc}
  \prod_{i=0}^{d-1}p(\C_{i+1}\rightarrow\C_i),
\end{equation*}
where the global acceptance probability is
\begin{equation*}
  p'_\text{acc}
  =m\bigl(\DE(\C_\text{E}\rightarrow\C_d)
  +\DE(\C_0\rightarrow\C_\text{S})\bigr).
\end{equation*}
When $\C_i=\C_{i+1}$,
\begin{equation*}
  \frac{p(\C_i\rightarrow\C_{i+1})}{p(\C_{i+1}\rightarrow\C_i)}=1
  =\exp\left(-\frac{\DE_{i\rightarrow i+1}}{k_BT}\right),
\end{equation*}
when $\C_i\neq\C_{i+1}$,
\begin{equation*}
  \frac{p(\C_i\rightarrow\C_{i+1})}{p(\C_{i+1}\rightarrow\C_i)}
  =\frac{m(\DE_{i\rightarrow i+1})}{m(\DE_{{i+1}\rightarrow i})}
  =\exp\left(-\frac{\DE_{i\rightarrow i+1}}{k_BT}\right),
\end{equation*}
and
\begin{equation*}
  \frac{p_\text{acc}}{p'_\text{acc}}
  =\exp\left(-\frac{\DE(\C_\text{S}\rightarrow\C_0)
    +\DE(\C_d\rightarrow\C_\text{E})}{k_BT}\right),
\end{equation*}
so that
\begin{flalign*}
  \frac{p(\C_\text{S}\rightarrow\C_\text{E})}
       {p(\C_\text{E}\rightarrow\C_\text{S})}
  &= \exp\left(-\frac{\DE_{\text{S}\rightarrow0}
    +\overset{d-1}{\underset{i=0}{\sum}}\DE_{i\rightarrow i+1}
    +\DE_{d\rightarrow\text{E}}}{k_BT}\right) \\
  &= \exp\left(-\frac{\mathcal{E}(\C_\text{E})
         -\mathcal{E}(\C_\text{S})}{k_BT}\right).
\end{flalign*}

\begin{figure}[hbt]
  \includegraphics[bb=0 0 226 163,clip=true,width=\colwid]{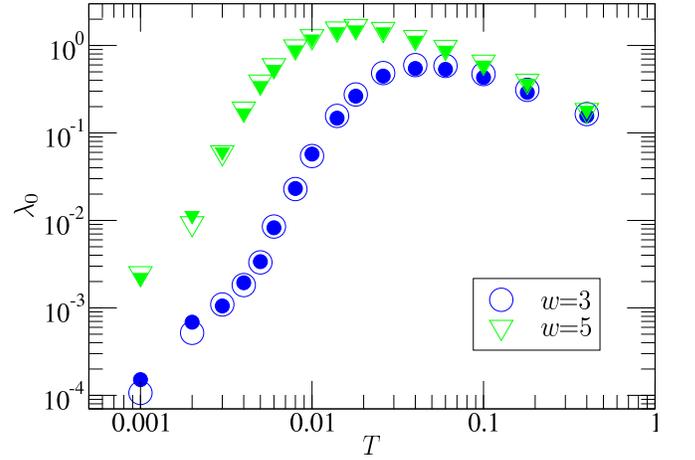}
  \caption{\label{fig:LRPM}
    Mobility of temporarily spread charges versus temperature.  OCP,
    $N=2$, $L=15$.  Open symbols, one trial move per update. Filled
    symbols, $d=15$ trial moves per update.}
\end{figure}

To check that the mobility is not changed when these long-ranged
particle moves are used, we simulated our test system with them, fixing
$d=15$. Regarding time units, one such update amounts to $d$ elementary
Monte Carlo steps. On Fig.~\ref{fig:LRPM} we find that the mobility
of charges is very little affected by the use of long distance particle
updates. However CPU cost is reduced, as is shown in next section.

\begin{table*}[t]
\newcommand{\esp}{\phantom{$\mathstrut\times10^{-4}$}}
\newcommand{\startline}[1]{\multicolumn{1}{|cI}{#1}}
\newcolumntype{I}{!{\vrule width 3\arrayrulewidth}}
\newlength\celldep\newlength\cellhei
\settodepth{\celldep}{\strut}
\settoheight{\cellhei}{\large\strut}
\newcolumntype{R}{>{\vrule width 0pt height \cellhei depth \celldep}r}
\newcolumntype{C}{>{\vrule width 0pt height \cellhei depth \celldep}c}
\begin{tabular}{lIR|r|dIr|r|dIr|r|dI}
  \cline{2-10} & \multicolumn{3}{CI}{Permanent spreading}
  & \multicolumn{3}{@{\hspace*{\stretch{1}}}m{4cm}@{\hspace*{\stretch{1}}}I}%
        {\centering Temporary spreading,\par precomputed current}
  & \multicolumn{3}{@{\hspace*{\stretch{1}}}m{4cm}@{\hspace*{\stretch{1}}}I}%
        {\centering Temporary spreading,\par long-ranged particle moves} \\
  \cline{2-10} & \multicolumn{1}{C|}{$\lambda_0~[\text{(VS)}^{-1}]$}
               & \multicolumn{1}{c|}{$t_\text{CPU}~[\text{s}]$}
               & \multicolumn{1}{cI}{efficiency~[s$^{-1}$]}
               & \multicolumn{1}{c|}{$\lambda_0~[\text{(VS)}^{-1}]$}
               & \multicolumn{1}{c|}{$t_\text{CPU}~[\text{s}]$}
               & \multicolumn{1}{cI}{efficiency~[s$^{-1}$]}
               & \multicolumn{1}{c|}{$\lambda_0~[\text{VS}^{-1}]$}
               & \multicolumn{1}{c|}{$t_\text{CPU}~[\text{s}]$}
               & \multicolumn{1}{cI}{efficiency~[s$^{-1}$]} \\
  \noalign{\hrule height 3\arrayrulewidth}
  \hline\startline{$w=1$}&   0\esp          &   71 &   0 &
                           0\esp          &   71 &   0 &
                           0\esp          &   71 &   0 \\

  \hline\startline{$w=3$}& $1.3\times10^{-3}$ &  160 &  0.49 &
                           $5\times10^{-2}$ &  592 &  5    &
                           $6\times10^{-2}$ &  233 & 15    \\

  \hline\startline{$w=5$}& $1.4\times10^{-2}$ &  582 &  1.4 &
                         1.3\esp          & 2372 & 33   &
                         1.2\esp          &  912 & 79   \\
  \noalign{\hrule height 3\arrayrulewidth}
\end{tabular}
  \caption{\label{table} Comparison of the various algorithms presented.
    $\lambda_0$, rate at which field configurations decorrelate, in
    simulation units.  $t_\text{CPU}$, duration of the 60~000~VS
    simulation.  Efficiency, real rate at which configurations
    decorrelate, given by $\lambda_0\times60~000/t_\text{CPU}$.  $w=1$
    stands for the single-link field update, displayed for execution
    time comparisons.}
\end{table*}

\section{Optimization\label{sec:optim}}

In Secs.~\ref{sec:spr} and \ref{sec:flum}, we presented several
ways of updating the electric field during charge motion. We also
measured the mobility of charges. However, the rates $\lambda_0$ have
been computed in simulation units; time is expressed in Monte Carlo
trials. As a function of their complexities, the different kinds of
update require different computational effort. In order to choose the
best parameters for a simulation we should express the efficiency of
the various versions of the algorithm in terms of CPU time.

We simulated $N=2$ mobile charges in a box of size $L=15$.  $T=0.01$
when charges are pointlike (temporary spreading) and $T=0.01/w$ when
spread. This set of parameters is representative for simulating a
monovalent ion in water.  At each elementary Monte Carlo step (MCS),
we try a particle move with probability $p_1=50\%$, and a plaquette
update with probability $p_2=50\%$.  We define a ``volume sweep''
(VS) $1~\text{VS}=L^3~\text{MCS}$.  $60~000~\text{VS}\gtrsim
2\times 10^8~\text{MCS}$ are performed after equilibration.  Temporary
spreading was implemented in both ways of Sec.~\ref{sec:flum}: first,
with steps 1 to 3 of Fig.~\ref{fig:floumpf} summed up and stored in a
single lookup table; second, with multiple steps 2 between each
spreading-and-recondensing pair of events.

In Table~\ref{table} we compare the efficiency of the various updates
introduced in this paper.  We used a Pentium~4 at 2.6~GHz; our C++
code was compiled with an Intel compiler. We conclude that the most
efficient field update is the temporary spreading of charges on $w=5$
cubes. At $T=0.01$, the mobility reached with $w=5$ is close to the
maximum possible value: $D\approx 0.15~a^2\,(\text{sweep})^{-1}$ is rather
close to saturation.
We thus do not expect benefit from further spreading of charges
($w\geqslant 6$).  As noted previously, both versions of temporary
spreading yield almost the same mobility.  The difference between the
two is CPU time: long-ranged particle moves lead to a faster algorithm
thanks to fewer spreading and recondensing steps.  This version should
thus be used for free charges.

Finally, we have checked that our results remain valid for higher
densities.  We applied our optimal solution (temporary spreading over
$5^3$ sites) to simulate OCPs containing $N=14$, $34$, and $336$
positive charges, which, respectively, corresponds to number densities
$n\approx 0.4\%$, $1\%$, and $10\%$.

\begin{figure}[hbt]
  \includegraphics[bb=0 0 226 163,clip=true,width=\colwid]{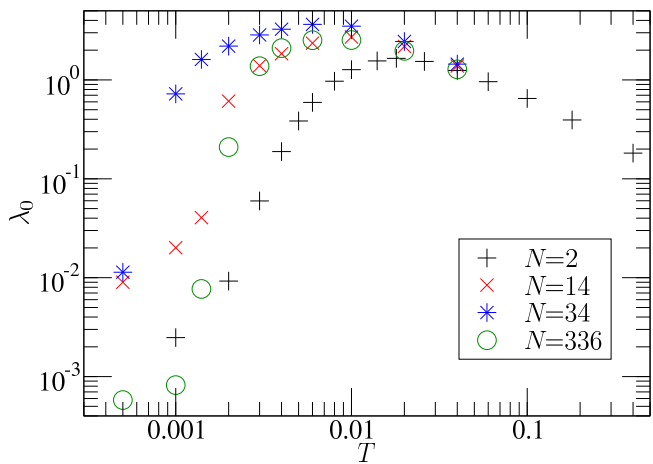}
  \caption{\label{fig:f_dens} Mobility of charges temporarily spread on
    $w=5$ cubes, for an OCP at various densities ($n=N/L^3$ with $L=15$).}
\end{figure}

In Sec.~\ref{sec:mobility}, we calculated a relationship according
to which $\lambda_0\propto n\mu$. This was for $\lambda_0$ in
physically relevant units of time, like particle sweeps (PS): the
effects of each charge add up, hence the factor of $n$. Here we
measure time in volume sweeps (VS), and work at constant numerical
effort, split amongst particles: the more charges there are, the fewer
trials each one does. $1~\text{VS}=0.5/n~\text{PS}$, so that
$\lambda_0[\text{in (VS)}^{-1}]
=0.5\lambda_0[\text{in (PS)}^{-1}]/n$ is directly proportional to
$\mu$.  Plotting mobility against temperature in
Fig.~\ref{fig:f_dens} we find that lowest mobility is found at the
lowest density;
at high density $\mu$ decreases, possibly because of steric hindrance,
but remains greater than when $N=2$.  Thus using our algorithm is
always at least as efficient as displayed in Table~\ref{table}.

\section{Conclusion}

The original version of the local Monte Carlo algorithm suffers from
two problems at low temperature. First the acceptance rate becomes
low due to a energy barrier for particle motion. A more serious
problem is that the mobility falls even faster than the acceptance
rate.  We  understand this fall in mobility by considering the
tension of the strings left behind particles as they move. The
different scaling of the acceptance rate and the mobility with the
spreading parameter $w$ is a clear demonstration that two different
mechanisms are important in limiting particle motion.

Simple modifications to the algorithm reduce the energy barrier for
single particle moves, but also the string tension.  The algorithm is
then suitable for simulation of lattice models of Coulomb interacting
particles.  Examples include the restricted primitive model for
electrolytes, or lattice models of polyelectrolytes.

Combination of the update methods used in this article with the worm
update for the transverse field due to Alet and Sørensen \cite{alet}
can lead to efficient codes for the simulation of charge systems at
high dilution: Consider a set of $N$ charges in a simulation box of
size $L$. It takes a computational effort of order $NL^2$ for these
particles to diffuse the system size.  We have already shown
\cite{StatPhys} that the $2L^3$ tranverse degrees of freedom of the
lattice can be integrated over in $\ord{1}$ sweeps of the worm
algorithm with an effort scaling as $L^3$. One can thus equilibrate a
dilute system of mobile charges with a computer effort which scales as
$ (NL^2 +L^3) $.

The time moving the particles dominates the time needed for the
electrostatic integration if $N>L$, or if the density $n>1/L^2$; when
$L$ is large the algorithm remains efficient even for \emph{very
  dilute} charges.  It is thus well suited to the study of
heterogenous systems such as surfaces and polyelectrolytes.

\bibliography{coulomb.bib}

\end{document}